\documentclass[titlepage, 12pt]{article}

\usepackage[utf8]{inputenc}
\usepackage{amsmath}
\usepackage{graphicx}
\usepackage{tikz}
\usepackage{setspace}
\usepackage{xcolor}
\usepackage{colortbl}
\usepackage{accents}
\usepackage{array}
\usepackage{upquote,textcomp}
\usepackage{parskip} 
\usepackage{pbox}
\usepackage{float}
\usepackage{tabularx,booktabs, multirow}
\usepackage[english]{babel}
\usepackage[autostyle, english = american]{csquotes}
\usepackage[normalem]{ulem}
\usepackage{subcaption, caption}
\usepackage[backend=biber,style=authoryear,citestyle=authoryear, maxbibnames=99, maxcitenames=2]{biblatex}
\usepackage{authblk}
\usepackage{soul}
\usepackage{rotating}
\usepackage{floatpag}

\MakeOuterQuote{"}

\DeclareNameAlias{sortname}{family-given}

\DeclareLabeldate[online]{%
  \field{date}
  \field{year}
  \field{eventdate}
  \field{origdate}
  \field{urldate}
}

\newcommand{\blind}{0}
\def\spacingset#1{\renewcommand{\baselinestretch}%
{#1}\small\normalsize} \spacingset{1}

\begin{document}

\if0\blind
{
\title{The Importance of Discussing Assumptions When Teaching Bootstrapping}

\author{%
Njesa Totty\thanks{Assistant Professor, Department of Mathematics, Framingham State University, Framingham, MA, 01701 (email: ntotty@framingham.edu)}
\and 
James Molyneux\thanks{Assistant Professor, Department of Statistics, Oregon State University, Corvallis, OR, 97331 (email: james.molyneux@oregonstate.edu)}
\and 
Claudio Fuentes\thanks{Associate Professor, Department of Statistics, Oregon State University, Corvallis, OR, 97331 (email: claudio.fuentes@oregonstate.edu)}
}

\maketitle
} \fi

\if1\blind
{
\title{The Importance of Discussing Assumptions when Teaching Bootstrapping}

\author[]{Blinded Version}

\maketitle
} \fi


\date{}

\begin{abstract}




\noindent
Bootstrapping and other resampling methods are increasingly appearing in the textbooks and curricula of courses that introduce undergraduate students to statistical methods. In order to teach the bootstrap well, students and instructors need to be aware of the assumptions behind these intervals. In this article we discuss important assumptions about simple non-parametric bootstrap intervals and their corresponding hypothesis tests. We present simulations that instructors can use to help students understand some of the assumptions behind these methods. The simulations will be especially relevant to instructors who desire to increase accessibility for students from non-mathematical backgrounds, including those with math anxiety. 

\vspace{10pt}
\end{abstract}

\vfill

\newpage
\spacingset{1.45} 

\section{Introduction}

The bootstrap is a simulation-based method introduced by \textcite{bradleybootstrap} as a technique for estimating the standard deviation of a sample statistic. In its simplest form, the term \emph{bootstrap sampling} refers to the process of randomly sampling with replacement from the original sample. This process is taken to be analogous to sampling from the entire population. Thus, the bootstrap estimate of the standard error is always available regardless of the form of the original estimator \parencite{tibshirani1993introduction}. 

This article focuses on the simplest non-parametric version of the bootstrap often taught in undergraduate introductory statistics courses. However, more complex applications of the bootstrap do exist and these are also popularly applied. Examples include the parametric bootstrap \parencite[e.g.][]{tibshirani1993introduction}, Bayesian bootstrap \parencite[e.g.][]{rubin1981bayesian, efron2012bayesian}, smoothed bootstrap \parencite[e.g.][]{efron1982jackknife, silverman1987bootstrap}, wild bootstrap \parencite[e.g.][]{wu1986jackknife}, and many others \parencite[e.g.][]{diciccio1988review, lahiri2006bootstrap}. The bootstrap has also found use in a variety of diverse applications such as linear regression \parencite[see][]{eck2018bootstrapping, pelawa2021bootstrapping} and bootstrap aggregated neural networks \parencite[see][]{khaouane2017modeling, osuolale2018exergetic}. 

The growth of statistical computing has led to the bootstrap appearing more regularly in courses which introduce undergraduate students to statistical methods with
examples including courses taught at Stanford University\footnotemark, The Pennsylvania State University\footnotemark, Oregon State University\footnotemark, and Montana State University\footnotemark. Textbooks about, or that feature, the bootstrap thus range from the seminal graduate-level text by \textcite{tibshirani1993introduction} to upper-level \parencite[e.g.][]{chihara2022mathematical} and intro-level \parencite[e.g.][]{field2012discovering, moderndive, lock2020statistics} undergraduate texts.

\footnotetext[1]{STAT 191 - Introduction to Applied Statistics at Stanford University (\url{https://explorecourses.stanford.edu})}

\footnotetext[2]{STAT 200 - Elementary Statistics at The Pennsylvania State University (\url{https://online.stat.psu.edu/stat200/})}

\footnotetext[3]{STAT 351/352 - Introduction to Statistical Methods I \& II (\url{https://stat.oregonstate.edu/content/yearly-courses})}

\footnotetext[4]{STAT 216 - Introduction to Statistics at Montana State University (\url{https://math.montana.edu/courses/s216/}). See also \textcite{hildreth2018comparing}.}

Teaching the bootstrap can equip students with a very powerful tool and lay a solid foundation for teaching statistical thinking. Therefore the discussion of how to teach the bootstrap well is an important one to have. Pedagogical discussions about whether bootstrap methods should be taught, which bootstrap methods to teach, and how to teach them include \textcite{hesterberg2015teachers} and \textcite{hayden2019questionable}. 

With this article we aim to further contribute to the discussion in two ways. First, we have compiled key theoretical details about simple bootstrap-based inferential methods and we explain these in a manner that is geared towards instructors of introductory statistics courses. Given that the majority of introductory statistics courses are taught outside of statistics departments \parencite[e.g.][]{garfield2000best, moore2001undergraduate, hulsizer2009guide} we purpose to make our discussion accessible to instructors with academic backgrounds outside of statistics. Second, we give examples of ways in which instructors can use simulations to explain some of the assumptions behind these methods to students in an accessible manner. 

By assumptions, we mean the hypotheses under which the theoretical results about these intervals are derived - in particular those having to do with pivotal quantities and normality. We discuss the basic, percentile, and studentized intervals and the $t$-interval with bootstrap standard error. These will be defined in Section 3. We also discuss corresponding hypothesis tests based on inversion of these intervals. We choose this focus because these methods are often taught in undergraduate introductory statistics courses.

In Section 2, we discuss the benefits of teaching the bootstrap and some issues pertaining to the bootstrap, as found in the literature on statistics education. In Section 3, we explain the theoretical details of the bootstrap and clearly point out the assumptions behind these methods. In Section 4, we provide examples of simulations that instructors can use to discuss some of the assumptions of these methods with students in introductory statistics courses. Lastly, in Section 5 we provide concluding remarks.

\section{Benefits of Teaching Statistical Computing and Bootstrapping}

According to the \emph{Guidelines for Assessment in Statistics Education} (GAISE), students in introductory statistics courses should, ``Demonstrate an understanding of, and ability to use, basic ideas of statistical inference, both hypothesis tests and interval estimation, in a variety of settings.'' \parencite[p. 8]{gaise} This implies that students should be able to recognize when a particular statistical method detracts from the quality of their analysis. Furthermore, upon realizing this, they should be able to pull an alternative method from their knowledge-base and apply it appropriately. 

Depending on learning goals and student backgrounds, including topics that incorporate statistical computing in a course, such as resampling, randomization, or simulation, can help students achieve these objectives. This idea resounds from an original and insightful discussion by \textcite{cobb2007introductory} which encourages instructors to center their courses on the logic of inference and emphasize differences between a model assumed for data and reality. Additonally, \textcite{wood2005role} notes that through general simulation methods students are able to ``actively and intelligently'' apply the methods they are taught to solve problems of current concern. 

Results of an early study by \textcite{tintle2012retention} found that the use of a randomization-based curriculum led to a higher retention of concepts after four months than using the consensus curriculum based on \textcite{agresti2018statistics}. More recently, simulation methods were incorporated by \textcite{learnsci} in their discussed ``practicing connections'' approach to building an introductory statistics course. Their approach was found to make students capable of applying previously learned material in new and more sophisticated contexts.

Regarding student preparedness for a variety of scenarios \textcite{howington2017teach} argues that introductory statistics courses often encourage use of the median as a measure of center when data are skewed but inferential methods are rarely taught for the median. They suggest a variety of methods for teaching confidence intervals on the median, including the bootstrap. The use of simulation-based inferential methods was also discussed by \textcite{gehrke2021statistics}. Their incorporation of methods like the bootstrap in an improved curriculum helped students to more clearly explain $p$-values and confidence intervals and to understand the limitations of statistics as it pertains to describing the real world. 

Additionally, \textcite{lock2008introducing} note that students' understanding of confidence intervals and statistical inference rests greatly on their understanding of sampling distributions. They note that teaching the bootstrap can allow students to make inference on non-conventional parameters and that their understanding of concepts like sampling distributions can be fortified through the use of simulations and bootstrapping. 

The idea that statistical computing should be taught more in order to better equip students for present-day workforce expectations undercurrents much of the literature on statistics in the undergraduate curriculum. Besides the aforementioned literature, various articles in the collection complied by \textcite{horton2015teaching} express this idea. Technically, the more statistical methods a student is introduced to, the better equipped they should be to meet the GAISE guideline discussed earlier and to tackle real-world data challenges. In reality, as students learn more statistical methods, discerning which one is appropriate to use becomes harder, especially if students are not clearly taught how to check whether a method is appropriate for their data. 

For example, many incorrect or unfounded claims have been made about simple bootstrap intervals, making it hard to know when their use is appropriate. \textcite{hayden2019questionable} investigated these claims in greater detail for the percentile bootstrap interval and the $t$-interval with bootstrap standard error which we will define in Section 3. They discuss the falsehood of the claim that these bootstrap intervals have fewer or no underlying assumptions than their traditional counterparts. They also found that these bootstrap intervals do not actually perform better when normality and large sample size conditions are not met. Their supposed simplicity was said to be the result of a failure to communicate their assumptions as clearly as those of the traditional methods. Introducing these intervals to students before appropriate scenarios for their use are better established was discouraged. \textcite{hesterberg2015teachers} discusses issues with the percentile as well as the basic bootstrap intervals and finds the studentized bootstrap interval, also called the bootstrap $t$-interval, to be preferable. 

There are obvious benefits to teaching the bootstrap in introductory statistics courses, but there are also important details and issues surrounding these methods that instructors and students should be aware of. Communicating this information to students, and specifically doing so in an accessible manner, can ensure that all students get an equal chance of reaping the benefits of these methods. In the next section we discuss these details and issues in a theoretical manner. This discussion will benefit the instructor who desires to understand more about the theory behind these methods. However, such a discussion will not be accessible to the typical introductory statistics student. Therefore, in Section 4 we provide example simulations that can be used to explain some of these assumptions to students in a more accessible manner.

\section{General Assumptions for Simple Applications of Bootstrapping}

In order to make an inference on a population parameter, $\theta$, we begin by taking an independent and identically distributed (iid) sample of size $N$, $\textbf{x} = (x_1, x_2, \ldots,x_N)$, from the population of interest. Let $\hat\theta(\textbf{x})$ denote an estimate for $\theta$ based on the original observed data. If it is based on the not yet observed data we use $\hat\theta(\textbf{X})$, where $\textbf{X}$ denotes the unobserved data vector. This sample should be taken in such a way that it captures most of the characteristics in the population that influence $\theta$. If $F$ is the population distribution of $X$, then the empirical distribution, $\hat{F}$, should approximate $F$ well. 

The estimate, $\hat\theta(\textbf{x})$, should summarize the information about $\theta$ that is contained in the observed data. For example, if $\theta$ is the population mean $\mu$, then $\hat\theta(\textbf{x})$ may be the observed sample mean $\bar{x}$. We often desire to gather more information than that contained in $\hat\theta(\textbf{x})$ alone. Options for achieving this include confidence intervals and hypothesis testing. Many methods exist for constructing confidence intervals and hypothesis testing, such as $t$-methods and jackknife-, bootstrap-, or permutation-based approaches. 

The concept of bootstrapping through simple random resampling is as follows: Obtain $B$ samples $\bf{x}^*_1, \bf{x}^*_2, \ldots, \bf{x}^*_B$, each of size $N$, by sampling with replacement from the original sample $\bf{x}$, and calculate their corresponding statistics $\hat\theta(\bf{x}^*_1), \hat\theta(\bf{x}^*_2), \ldots, \hat\theta(\bf{x}^*_B)$. These bootstrap statistics make up the bootstrap distribution. 

\subsection{Interval estimation}

The bootstrap distribution can be used as an estimate of the sampling distribution, which provides a means for quantifying the uncertainty in an estimate. The basic, percentile, and studentized bootstrap intervals each use the bootstrap distribution in this manner but have different underlying assumptions, most of which pertain to the shifted or studentized sampling distribution. The details we discuss next will be helpful for readers who desire to become familiar with these bootstrap intervals as they are presented by \textcite{davison1997bootstrap} and \textcite{tibshirani1993introduction}. Our explanation is not exhaustive, however, so readers who desire a more in-depth understanding of these methods and their assumptions should consult those texts directly, as well as \textcite{hallboot} and \textcite{shao2012jackknife}. 

\subsubsection{The Basic Interval}

Let $0 < \alpha < 1$ denote the significance level or desired Type I error rate. In order to construct a $(1 - \alpha)100\%$ confidence interval for $\theta$, we may employ a \emph{pivotal quantity} - a quantity whose distribution does not depend on any unknown parameters. When this quantity is a function of the parameter and estimate, a transformation of the quantiles of its distribution can be used to construct confidence intervals for the parameter.

Suppose that $\hat\theta(\textbf{X}) - \theta$ is a pivotal quantity. We denote its $\alpha/2$ and $1-\alpha/2$ quantiles as $a_{\alpha/2}$ and $a_{1-\alpha/2}$, respectively. If $a_{\alpha/2}$ and $a_{1-\alpha/2}$ are known, then $$1-\alpha = P(a_{\alpha/2} \leq \hat\theta(\textbf{X}) - \theta \leq  a_{1-\alpha/2}) 
    = P(\hat\theta(\textbf{X})-a_{1 - \alpha/2} \leq \theta \leq \hat\theta(\textbf{X})-a_{\alpha/2})$$ 
and a $(1-\alpha)100\%$ equi-tailed interval for $\theta$, provided the expression exists, is 
\begin{equation} \label{eq:1}
\left(\hat\theta(\textbf{x}) - a_{1 -\alpha/2}, \hat\theta(\textbf{x}) - a_{\alpha/2}\right).
\end{equation}

In a similar manner, $$1 - \alpha = P(\hat\theta(\textbf{X}) - \theta \geq a_{\alpha})  = P(\theta \leq \hat\theta(\textbf{X})-a_{\alpha})$$ and $$1 - \alpha = P(\hat\theta(\textbf{X}) - \theta \leq a_{1-\alpha}) = P(\theta \geq \hat\theta(\textbf{X})-a_{1-\alpha})$$ and $(1-\alpha)100\%$ one-sided intervals for $\theta$, provided the expressions exist, are 
\begin{equation} \label{eq:2}
\left(-\infty, \hat\theta(\textbf{x}) - a_{\alpha}\right) \quad \text{ and } \quad \left(\hat\theta(\textbf{x}) - a_{1 -\alpha}, +\infty\right).
\end{equation}

When the distribution of $\hat\theta(\textbf{X}) - \theta$ is unknown, the problem becomes one of estimating its quantiles. The basic bootstrap interval is obtained by estimating the $p$ quantile $a_p$ with the $(B+1)p$-th smallest value of the distribution of $\hat\theta(\textbf{x}^*) - \hat\theta(\textbf{x})$. For example, if $\alpha = 0.05$ and $B = 999$ then 
$$(B+1)(\alpha/2) = (999+1)0.025 
    = 1000*0.025 
    = 25$$ 
and similarly, $(B+1)(1 - \alpha/2) = 975$. 
Thus, the 25th smallest and 975th smallest values of the distribution of $\hat\theta(\textbf{x}^*) - \hat\theta(\textbf{x})$, denoted as $a^*_{(25)}$ and $a^*_{(975)}$, would be used to estimate $a_{0.025}$ and $a_{0.975}$, respectively. Note that the 25th smallest value is less than the 975th smallest value so the upper bound will be greater than the lower bound since we subtract off a smaller number. 

In this article, we assume that $(B+1)\alpha$ and $(B+1)(1-\alpha)$ are integers. If they are not, then the procedure outlined by \textcite{tibshirani1993introduction} can be used. Assuming $\alpha \leq 0.5$, define $k$ as the largest integer that is $\leq (B+1)\alpha$. Then the $\alpha$ and $1-\alpha$ quantiles are defined as the $k$-th largest and $(B+1-k)$-th largest values of the distribution of interest, respectively.

Using bootstrap estimates, the expressions in (\ref{eq:1}) and (\ref{eq:2}) become $$\left(\hat\theta(\textbf{x}) - a^*_{\left((B+1)(1 -\alpha/2)\right)},\hat\theta(\textbf{x}) - a^*_{\left((B+1)(\alpha/2)\right)}\right),$$ $$\left(-\infty, \hat\theta(\textbf{x}) - a^*_{((B+1)(\alpha))}\right), \textrm{ and } \left(\hat\theta(\textbf{x}) - a^*_{((B+1)(1 -\alpha))}, +\infty\right). $$ We denote the $p$ quantiles of the distributions of $\hat\theta(\textbf{X})$ and $\hat\theta(\textbf{x}^*)$ as $r_p$ and $r^*_p$, respectively. Order is preserved after shifting the distributions of these quantities, therefore
$$a_{((B+1)p)} = r_{((B+1)p)} - \theta
    \quad 
    \textrm{and} 
    \quad 
    a^*_{((B+1)p)} = r^*_{((B+1)p)} - \hat\theta(\textbf{x}).$$ 
The bounds of the two-sided interval can thus be simplified to 
$$\hat\theta(\textbf{x}) - a^*_{((B+1)(1 -\alpha/2))} = 2\hat\theta(\textbf{x}) - r^*_{((B+1)(1 -\alpha/2))} 
    \quad
    \textrm{and}
    \quad
    \hat\theta(\textbf{x}) - a^*_{((B+1)(\alpha/2))} = 2\hat\theta(\textbf{x}) - r^*_{((B+1)(\alpha/2))}.$$ 
The final form of the two-sided \emph{basic bootstrap interval} is then $$\left(2\hat\theta(\textbf{x}) - r^*_{((B + 1)(1 - \alpha/2))}, 2\hat\theta(\textbf{x}) - r^*_{((B + 1)(\alpha/2))}\right).$$ Similarly the one-sided basic bootstrap intervals are $$\left(-\infty, 2\hat\theta(\textbf{x}) - r^*_{((B + 1)(\alpha))}\right) \textrm{ and } \left(2\hat\theta(\textbf{x}) - r^*_{((B + 1)(1 - \alpha))}, +\infty\right).$$

The accuracy of this interval depends on how well the distribution of $\hat\theta(\textbf{x}^*) - \hat\theta(\textbf{x})$ conforms to that of $\hat\theta(\textbf{X}) - \theta$. If differences in the sampling distribution depend only on a location parameter then $\hat\theta(\textbf{X}) - \theta$ is actually a pivotal quantity and a good approximation can be expected. This assumption is rarely reasonable for most statistics discussed in an introductory statistics course though. Additionally, this interval is not \textit{range-preserving}. If there are any constraints on the value of $\theta$, the bounds of this interval may not meet these constraints. For example, one may obtain values below 0 or above 1 in an interval for the population proportion. We agree with \textcite{hesterberg2015teachers} that this interval should not be used and we will not include further discussion about it. We refer readers to \textcite{hesterberg2015teachers} for a discussion about its poor performance and reasons not to use it.

\subsubsection{The Percentile Interval}

Suppose that the distribution of $\hat\theta(\textbf{X})$ is asymptotically Normal with mean $\theta$ and variance ${\textrm{SE}(\hat\theta(\textbf{X}))}^2$, where $\textrm{SE}(\hat\theta(\textbf{X}))$ denotes the standard error of $\hat\theta(\textbf{X})$. This provides another option for estimating $a_{p}$. Namely, with ${\hat{\textrm{SE}}(\hat\theta(\textbf{x}))}*z_{p}$, where $z_p$ is the $p$ quantile of the standard normal distribution. For example, in the case of the sample mean we may use ${\hat{\textrm{SE}}(\hat\theta(\textbf{x}))} = s/\sqrt{n}$ where $s$ denotes the sample standard deviation. The use of standard normal quantiles with an estimated standard error produces what is termed the \textit{standard normal interval} \parencite[e.g.][]{tibshirani1993introduction}. This is given as  $$\bigg(\hat\theta - {\hat{\textrm{SE}}(\hat\theta(\textbf{x}))}*z_{1-\alpha/2}, \hat\theta - {\hat{\textrm{SE}}(\hat\theta(\textbf{x}))}*z_{\alpha/2}\bigg).$$ 

This interval can have poor performance since the estimated standard error adds more variability. An alternative is the \textit{percentile interval} introduced by \textcite{tibshirani1993introduction} which is given as $$(r^*_{\alpha/2}, r^*_{1-\alpha/2}),$$ where we again denote the $p$ quantile of the distribution of $\hat{\theta}(\textbf{x}^*)$ as $r^*_p$. 

The assumption behind the percentile interval is that there exists some monotone transformation $\hat\phi = m(\hat\theta(\textbf{X}))$ such that $\hat\phi \sim \text{Normal}(\phi, c^2)$ for all population distributions $F$, including the case $F = \hat{F}$, were $\phi = m(\theta)$, for some standard deviation $c$. Then, it holds that $$1 - \alpha = P\bigg(z_{\alpha/2} \leq \frac{\hat\phi - \phi}{c} \leq z_{1-\alpha/2}\bigg) = P(-\hat\phi + z_{\alpha/2}\cdot c \leq -\phi \leq -\hat\phi + z_{1-\alpha/2} \cdot c)=$$ $$P(m^{-1}(\hat\phi - z_{1-\alpha/2}\cdot c) \leq \theta \leq m^{-1}(\hat\phi - z_{\alpha/2} \cdot c)).$$ 

Since the assumption holds for $F = \hat{F}$ it is also the case that $\hat\phi^* \sim \text{Normal}(\hat\phi, c^2)$, where $\hat\phi^* = m(\hat\theta(\textbf{x}^*))$. Therefore, $$1 - \alpha = P_*\bigg(z_{\alpha/2} \leq \frac{\hat\phi^* - \hat\phi}{c} \leq z_{1-\alpha/2}\bigg) = P_*(\hat\phi + z_{\alpha/2}\cdot c \leq \hat\phi^* \leq \hat\phi + z_{1-\alpha/2} \cdot c)=$$ $$P_*(m^{-1}(\hat\phi - z_{1-\alpha/2}\cdot c) \leq \hat\theta(\textbf{x}^*) \leq m^{-1}(\hat\phi - z_{\alpha/2} \cdot c)).$$ Moreover we see that $r^*_{\alpha/2} = m^{-1}(\hat\phi - z_{1-\alpha/2}\cdot c)$ and $r^*_{1 - \alpha/2} = m^{-1}(\hat\phi - z_{\alpha/2} \cdot c))$. Therefore the percentile interval agrees with the standard normal interval applied to the appropriate transformation of $\theta$. That is, the transformation that causes the assumptions of the standard normal interval to actually hold. However, the appropriate transformation does not need to be known to construct the percentile interval, making it superior to the standard normal interval.

For a finite number of bootstrap replications the two-sided percentile interval is $$(r^*_{(B+1)(\alpha/2)}, r^*_{(B+1)(1-\alpha/2)}).$$ Similar derivations can also be used to obtain the one-sided versions: $$(-\infty, r^*_{(B+1)(1-\alpha)}) \textrm{ and } (r^*_{(B+1)(\alpha)}, +\infty).$$ When the assumptions of the standard normal interval are met, the percentile interval will agree with it. When the standard normal interval would be correct for a certain transformation, the percentile interval agrees with the results of the standard normal interval applied under that transformation. There are many cases in which the assumption that such a transformation exists is quite reasonable. Such as when $\hat\theta(\textbf{X})$ is the sample mean, proportion, or a regression coefficient. In these and other cases where a central limit theorem applies the identity transformation suffices. 

Since its introduction the percentile interval has been interpreted in a pivotal framework \parencite[e.g.][]{hinkley1988bootstrap, shao2012jackknife}. If the distribution of $\hat\theta(\textbf{X}) - \theta$ is symmetric, then $-a_{1 - \alpha/2} = a_{\alpha/2}$ and $a_{1-\alpha/2} = -a_{\alpha/2}$. Therefore, we can rewrite (\ref{eq:1}) as
$$\left(\hat\theta(\textbf{x}) + a_{\alpha/2}, \hat\theta(\textbf{x}) + a_{1 - \alpha/2}\right).$$ 
Upon estimating these quantiles with the appropriate order statistics from the bootstrap distribution we obtain 
$$\left(\hat\theta(\textbf{x}) + a^*_{((B + 1)(\alpha/2))},
    \hat\theta(\textbf{x}) + a^*_{((B + 1)(1-\alpha/2))}\right).$$ 
Observe that $a^*_{((B + 1)p)} = r^*_{((B + 1)p)} - \hat\theta(\textbf{x})$, so instead we can write 
$$\hat\theta(\textbf{x}) + a^*_{((B + 1)(\alpha/2))} = r^*_{((B + 1)(\alpha/2))} 
    \quad 
    \textrm{and} 
    \quad 
    \hat\theta(\textbf{x}) + a^*_{((B + 1)(1-\alpha/2))} = r^*_{((B + 1)(1 - \alpha/2))}.$$ Hence we arrive at the
    same quantiles of the bootstrap distribution.
    
The simplicity of the percentile interval provides a pedagogical advantage. Students can easily verify if the method is appropriate by checking the bootstrap distribution for normality. The interval is also transformation-respecting and range-preserving. However, the nonparametric percentile interval has received criticism for its poor performance \parencite[e.g.][]{hinkley1988bootstrap, hesterberg2015teachers, hayden2019questionable}. It has also been noted that the percentile interval uses the ``wrong pivot backwards'' relative to the basic interval \parencite[e.g.][]{hallboot}. This is discussed by \textcite{tibshirani1993introduction} who state that neither the percentile nor basic intervals, ``work well in general''. However, they note that the percentile interval works better than the basic interval in practice. 

A suggested improvement to the percentile interval is the bias-corrected and accelerated percentile interval, which accounts for possible bias in $\hat\theta(\textbf{X})$. Its details are discussed in Chapter 14 of \textcite{tibshirani1993introduction}. These details are more intricate and complex than those of the percentile and basic interval and, depending on the students' mathematical backgrounds, they may be outside of the scope of an undergraduate introductory statistical methods course.

\subsubsection{$t$-Interval with Bootstrap Standard Error}

The standard normal interval uses quantiles from the standard normal distribution which does not account for estimating the standard error of $\hat\theta(\textbf{X})$. A better approximation can be obtained by using the quantiles of a $t_{n - 1}$ distribution. As the sample size increases the $p$ quantile of the $t$ distribution decreases to that of the standard normal distribution.  When $\hat\theta(\textbf{X})$ is the sample mean and the data are normally distributed the approximation is exact. The approximation improves as the sample size increases for non-normal data, relative to the skewness of the underlying population.

The $t$\textit{-interval with bootstrap standard error} follows the form of the usual $t$-interval but uses the bootstrap standard error as an estimate for ${\hat{\textrm{SE}}(\hat\theta(\textbf{X}))}$. The \emph{plug-in principle} discussed by \textcite{tibshirani1993introduction} is used to estimate the standard error of $\hat\theta(\textbf{X})$ with the square root of $${\hat{\textrm{SE}}_B(\hat\theta(\textbf{x}))}^2 = \frac{1}{B-1}\displaystyle\sum_{i=1}^B\bigg(\hat\theta(\textbf{x}_i^*) - \bar{\hat\theta}(\textbf{x}^*_{(\cdot)})\bigg)^2,$$ 
where $\bar{\hat\theta}(\textbf{x}^*_{(\cdot)})$ denotes the mean of the bootstrap sample statistics.

Let $t_{p, n-1}$ be the $p$ quantile of a $t_{n-1}$ distribution. The two-sided and one-sided intervals are then $$\bigg(\hat\theta(\textbf{x}) - \hat{\textrm{SE}}_B(\hat\theta(\textbf{x}))*t_{1-\alpha/2, n-1}, \hat\theta(\textbf{x}) - \hat{\textrm{SE}}_B(\hat\theta(\textbf{x}))*t_{\alpha/2, n-1}\bigg), $$ $$\bigg(-\infty, \hat\theta(\textbf{x}) - \hat{\textrm{SE}}_B(\hat\theta(\textbf{x}))*t_{\alpha, n-1}\bigg) \text{ and }  \bigg(\hat\theta(\textbf{x}) - \hat{\textrm{SE}}_B(\hat\theta(\textbf{x}))*t_{1-\alpha, n-1}, +\infty\bigg).$$

When $\hat\theta(\textbf{X})$ is not the sample mean the bootstrap standard error can be quite helpful for constructing an interval. However, use of the $t$ distribution only corrects for estimating the standard error, not for skewness in the underlying population and other issues that can arise when $\hat\theta(\textbf{X})$ is not the sample mean. 

\subsubsection{The Studentized Interval (also called the bootstrap-$t$ interval)}

The \textit{studentized interval}, also known as the \textit{bootstrap} $t$\textit{-interval}, is motivated by the form of the $t$-interval but it is useful outside of inference for the mean. Rather than using a $z$- or $t$-table, the studentized bootstrap interval uses ``bootstrap tables'' to approximate the critical value. These tables are fit for the specific data set observed. This adjusts for skewness in the underlying population and other errors that can arise when $\hat\theta({\textbf{X}})$ is not the sample mean. 

The value $t_{p, (n-1)}$ is estimated with the $(B+1)p$-th smallest value of the distribution of $z^* = \left(\hat\theta(\textbf{x}^*) - \hat\theta(\textbf{x})\right)/{\hat{\textrm{SE}}(\hat\theta(\textbf{x}^*))},$ where ${\hat{\textrm{SE}}(\hat\theta(\textbf{x}^*))}$ is a bootstrap estimate of the standard error of $\hat\theta(\textbf{x}^*)$. Substituting these bootstrap estimates leads to the following two-sided and one-sided intervals $$\left(\hat\theta(\textbf{x}) - {\hat{\textrm{SE}}_B(\hat\theta(\textbf{x}))}*z^*_{((B+1)(1-\alpha/2))},~~ \hat\theta(\textbf{x}) - {\hat{\textrm{SE}}_B(\hat\theta(\textbf{x}))}*z^*_{((B+1)(\alpha/2))}\right),$$ $$\left(-\infty, ~~ \hat\theta(\textbf{x}) - {\hat{\textrm{SE}}_B(\hat\theta(\textbf{x}))}*z^*_{((B+1)(\alpha/2))}\right) \text{ and } \left(\hat\theta(\textbf{x}) - {\hat{\textrm{SE}}_B(\hat\theta(\textbf{x}))}*z^*_{((B+1)(1-\alpha/2))}, ~~ \infty \right).$$ 

The studentized bootstrap interval requires estimates for the standard errors of $\hat\theta(\textbf{X})$ and $\hat\theta(\textbf{X}^*)$. In some cases, such as the sample mean, there exists a simple formula for the standard error that students can easily understand. We can obtain ${\hat{\textrm{SE}}(\hat\theta(\textbf{x}))} = s/\sqrt{n}$ from the original sample and ${\hat{\textrm{SE}}(\hat\theta(\textbf{x}^*))} = s^*/\sqrt{n}$ from each bootstrap sample. In other cases the \emph{plug-in principle} discussed by \textcite{tibshirani1993introduction} can be used to estimate the standard error of $\hat\theta(\textbf{X})$ with ${\hat{\textrm{SE}}_B(\hat\theta(\textbf{x}))}$. 

An iterative bootstrap method can be used to estimate the standard error of $\hat\theta(\textbf{x}^*)$ when no formula is available. In this method one obtains $M$ second-level bootstrap samples from \emph{each} of the $B$ original bootstrap samples. For each of these second-level bootstrap samples, $M$ statistics are then calculated and denoted as $\hat\theta(\textbf{x}^*_{i,j})$ for $i = 1, \ldots, B$ and $j = 1, \ldots,M$. From these we calculate the bootstrap estimate of standard error for the $i^{th}$ bootstrap sample as the square root of $${\hat{\textrm{SE}}_M(\hat\theta(\textbf{x}_i^*))}^2 = \frac{1}{M-1}\displaystyle\sum_{j=1}^M\left(\hat\theta(\textbf{x}^*_{i,j}) - \bar{\hat\theta}(\textbf{x}^*_{i(\cdot)})\right)^2,$$ where $\bar{\hat\theta}(\textbf{x}^*_{i(\cdot)})$ now represents the mean of the second-level bootstrap sample statistics. This method is useful for teaching inference when the form of the standard error is unknown, but it may be more difficult for introductory statistics students to grasp than formula-based methods and computations can take longer. 

For estimating quantiles, using $B \geq 999$ is encouraged. Though \textcite{tibshirani1993introduction} and \textcite{davison1997bootstrap} suggest values for $M$ ranging from $10$ to $50$, using $M \geq 100$ is now feasible given recent developments in computing power. As with the basic bootstrap interval, the accuracy of this interval depends on whether the distribution of $(\hat\theta(\textbf{X}) - \theta)/{\hat{\textrm{SE}}(\hat\theta(\textbf{X}))}$ is indeed pivotal. The quantity $z^*$ can be plotted against theoretical quantiles from a $t$ distribution to verify the pivotal assumptions. \textcite{hesterberg2015teachers} examples this and also found the studentized interval to be preferable to a variety of others studied. It is noted by \textcite{tibshirani1993introduction} that the results of the studentized bootstrap interval can be largely influenced by outliers in the data. They also warn that the studentized bootstrap interval works best for variance-stabilized parameters and that it is especially applicable to location statistics.

\subsubsection{Two-Sample Intervals Based on Simple Bootstrap}

Each of the intervals discussed above can also be extended to the two-sample case. This is useful when we desire to make inference on some population parameter, say $\theta$, that compares two populations. We begin by independently drawing two iid samples $\textbf{x} = (x_1, x_2, \ldots,x_{N_1})$ and $\textbf{z} = (z_1, z_2, \ldots,z_{N_2})$ from the populations of interest. Using these two samples we calculate a statistic $\hat\theta(\textbf{x}, \textbf{z})$ to estimate $\theta$. For example, if we are interested in the difference in population medians then $\theta = M_X - M_Z$ is the population parameter, where $M$ represents the population median. Here we use non-bolded capital letters to denote the population that the parameter belongs to while for unobserved data we will still use bolded capital letters. The observed estimate may be $\hat\theta(\textbf{x}, \textbf{z}) = m_\textbf{x} - m_\textbf{z}$, where $m$ represents the sample median. Though we denote the statistic as $\hat\theta$, it does not necessarily have to be the plug-in estimate of $\theta$.

In order to calculate bootstrap sample statistics we take bootstrap samples from each of our original samples separately and apply the functional form of $\hat\omega$ to them. Continuing with the example of the median, we take two bootstrap samples $\textbf{x}^*_i, \textbf{z}^*_i$ from the original samples and calculate $\hat\theta^*_i(\textbf{x}^*_i, \textbf{z}^*_i) = m_{\textbf{x}_i^*} - m_{\textbf{z}_i^*}$. Repeating this $B$ times, we obtain a bootstrap distribution which can then be used to construct bootstrap intervals for $\theta$ in a manner analogous to the one-sample case.

We must also calculate standard errors in order to use the $t$-interval with bootstrap standard error or studentized interval. When a formula exists for the standard error of $\hat\theta(\textbf{x}, \textbf{z})$ it can be used to calculate an estimated standard error for the original sample and each bootstrap sample. For example, if $\hat\theta(\textbf{x}, \textbf{z}) = \hat{p}_\textbf{x} - \hat{p}_\textbf{z}$, the difference in proportions, then a known formula for the standard error exists based on statistical theory. The estimates are then, $$\hat{\text{SE}}(\hat\theta(\textbf{x}, \textbf{z})) = \sqrt{\frac{\hat{p}_\textbf{x}(1 - \hat{p}_\textbf{x})}{N_1} + \frac{\hat{p}_\textbf{z}(1-\hat{p}_\textbf{z})}{N_2}}$$ and $$\hat{\text{SE}}(\hat\theta(\textbf{x}^*_i, \textbf{z}^*_i)) = \sqrt{\frac{\hat{p}_{\textbf{x}_i^*}(1 - \hat{p}_{\textbf{x}_i^*})}{N_1} + \frac{\hat{p}_{\textbf{z}_i^*}(1-\hat{p}_{\textbf{z}_i^*})}{N_2}}.$$ When a known formula does not exist, we may resort back to the bootstrap-based estimates introduced in the previous two sections. These we denote as ${\hat{\textrm{SE}}_B(\hat\theta(\textbf{x}, \textbf{z}))}$ and ${\hat{\textrm{SE}}_M(\hat\theta(\textbf{x}, \textbf{z}))}$.

We have provided a brief discussion on extensions to simple two-sample scenarios. For brevity we do not restate each of the intervals for the two-sample case. They and their assumptions are analogous to the one-sample intervals discussed earlier.

\subsection{Bootstrap-Based Hypothesis Tests}

The goal of hypothesis testing is to make an inference about some population parameter of interest, $\theta$, specifically in regards to whether or not there is sufficient evidence to indicate that the parameter is a value other than one which we hypothesize to be true. Many early manuscripts and textbooks \parencite[e.g.][]{beran1988prepivoting, hinkley1988bootstrap, tibshirani1993introduction,davison1997bootstrap} give guidance on bootstrap hypothesis testing and discuss possible approaches. 

In order to conduct a one-sample level-$\alpha$ hypothesis test of $H_0: \theta = \theta_0$ based on the simple bootstrap, two components must be obtained: (1) $t(\textbf{X})$, a test statistic, and (2) $\hat{T}_0$, an estimate of T, the distribution of $t(\textbf{X})$, under $H_0$. Given these two components we can calculate an approximate $p$-value as the proportion of test statistics that are as or more extreme than $t(\textbf{X})$ under $\hat{T}_0$. 

We briefly discuss some ideas presented by \textcite{tibshirani1993introduction} and \textcite{davison1997bootstrap} for obtaining the second component. Our discussion is brief since these hypothesis tests rely heavily on the intervals that we have just discussed in detail. Our goal is to point out the connection between simple bootstrap intervals and hypothesis tests. Readers who desire more details about bootstrap-based hypothesis tests are encouraged to read Chapter 4 of \textcite{davison1997bootstrap} and Chapters 12.3, 15, and 16 of \textcite{tibshirani1993introduction}.

\subsubsection{Inversion of Bootstrap Intervals}

In order to obtain $\hat{T}_0$ we may transform the original dataset and draw bootstrap samples from $\hat{F}_0$, an empirical distribution that obeys the null hypothesis \parencite[e.g.][]{tibshirani1993introduction}. However, \textcite{hesterberg2015teachers} notes that transforming the data can lead to negative values for data that must be positive and that this method is less accurate than permutation tests when they can be used. An alternative is to invert a bootstrap interval and simply reject values for $\theta_0$ that it does not contain. In some cases (See Chapter 16 of \textcite{tibshirani1993introduction}) this method produces the same results as transforming the data values.

If we desire to determine how plausible $\theta_0$ is, then we can construct the test statistic $t(\textbf{x}) = \big(\hat\theta(\textbf{x}) - \theta_0\big)/{\hat{\textrm{SE}}\big(\hat\theta(\textbf{x})\big)}$. Assuming that $t(\textbf{X}) = \big(\hat\theta(\textbf{X}) - \theta\big)/{\hat{\textrm{SE}}\big(\hat\theta(\textbf{X})\big)}$ is pivotal, then its distribution is the same regardless of $H_0$. Therefore we can approximate $T_0$, the distribution of $t(\textbf{x})$ under the null hypothesis, with the distribution of $t(\textbf{x}^*) = \big(\hat\theta(\textbf{x}^*) - \hat\theta(\textbf{x})\big)/{\hat{\textrm{SE}}\big(\hat\theta(\textbf{x}^*)\big)}$. To calculate $p$-values we determine which proportion of values $t(\textbf{x}^*)$ are as or more extreme than $t(\textbf{x})$, with respect to the alternative hypothesis.

However, note that if $\theta_0$ is contained in the studentized interval, then $$z^*_{((B+1)(\alpha/2))} < \frac{\big(\hat\theta(\textbf{x}) - \theta_0\big)}{{\hat{\textrm{SE}}\big(\hat\theta(\textbf{x})\big)}} < z^*_{((B+1)(1-\alpha/2))}.$$ Since the quantity in the center is just the observed test statistic with respect to $\theta_0$, containment of $\theta_0$ in the two-sided studentized interval implies that this test statistic will fall in the rejection region of the two-sided level-$\alpha$ hypothesis test based on $t(\textbf{x}^*)$. Therefore, performing this hypothesis test is equivalent to rejecting values of $\theta_0$ which are not contained in the studentized interval. A similar explanation shows the relationship between one-sided tests and intervals.

If instead we are willing to assume that $t(\textbf{X}) = \big(\hat\theta(\textbf{X}) - \theta\big)$ is pivotal, then we can denote our observed test statistic as $t(\textbf{x}) = \big(\hat\theta(\textbf{x}) - \theta_0\big)$ and estimate $T_0$ with the distribution of $t(\textbf{x}^*) = \big(\hat\theta(\textbf{x}^*) - \hat\theta(\textbf{x})\big)$ regardless of $H_0$. If $\theta_0$ is contained in the basic interval then, $$r^*_{(B+1)(\alpha/2)} - \hat\theta(\textbf{x}) < \hat\theta(\textbf{x}) - \theta_0 < r^*_{(B+1)(1-\alpha/2)} - \hat\theta(\textbf{x}).$$ The quantity in the center of the inequality is the observed test statistic and we see that values contained in the basic bootstrap interval, or the percentile bootstrap interval under the symmetry assumption, are in the rejection region of a test based on this quantity.

To perform two-sample bootstrap hypothesis tests one may invert the two-sample versions of these intervals. Inverting intervals is not unique to bootstrap hypothesis testing. The $t$-interval is based on the same underlying idea, with additional assumptions about the shape of the distribution of the test statistic.

\subsubsection{Calculating $p$-values}

Failing to reject values for $\theta_0$ that are contained in the bootstrap interval gives us an indication of which values are plausible for $\theta$. By calculating approximate p-values we can determine how plausible a certain value may be. For a one-sided lower alternative hypothesis $H_A: \theta < \theta_0$, \textcite{tibshirani1993introduction} suggest calculating the proportion of bootstrap test statistics that are less than or equal to the observed test statistic. This approximate $p$-value is defined as $$p = P^*\left(t(\textbf{x}^*) \leq t(\textbf{x})\right),$$ where the asterisk indicates that we use the distribution of the bootstrap test statistics generated under the null hypothesis to calculate the proportion. 

If the alternative hypothesis is one-sided upper, then $$p = P^*\left(t(\textbf{x}^*) \geq t(\textbf{x})\right)$$ and the approximate $p$-value is the proportion of bootstrap test statistics that are greater than or equal to the observed test statistic. If the alternative hypothesis is two-sided, then $$p = 2\times\min\left\{P^*\left(t(\textbf{x}^*) \leq t(\textbf{x})\right), P^*\left(t(\textbf{x}^*) \geq (\textbf{x})\right)\right\}.$$ In this case the approximate $p$-value is two times the smaller of the one-sided $p$-values which does not require symmetry in the distribution of bootstrap test statistics. In all cases, we reject $H_0$ if $p < \alpha$, where $0 < \alpha < 1$ is the desired significance level. 

\subsection{Summary}

In this section we have discussed the theoretical underpinnings of popular bootstrap intervals and how they can be inverted to perform hypothesis testing. Each of these methods relies on some assumption about the distribution of $\hat\theta(\textbf{X})$. The basic interval relies on the assumption that the distribution can be made approximately pivotal after shifting by $\theta$. The studentized intervals assumes the same but after additionally scaling by ${\hat{\textrm{SE}}(\hat\theta(\textbf{X}))}$. Meanwhile, the percentile interval relies on the assumption that a normalizing transformation of the sampling distribution exists. 

Whether the assumptions of these intervals are indeed reasonable depends on the parameter of interest and the underlying population data. For example, when $\hat\theta(\textbf{X})$ is the sample mean its distribution depends on a scale parameter and the assumption of the basic interval is quite unreasonable. These details are quite theoretical and explaining them in their full complexity will likely detract from the learning experience of introductory statistics students. Therefore, in the next section we discuss simulations that instructors can perform in-class with students in order to help them understand some of these assumptions. We aim to equip instructors of introductory statistics courses with tools to maximize their time spent teaching simple bootstrap methods.

\section{Using Simulations To Teach The Bootstrap}

In this section we discuss simulations that instructors and students can use to understand the assumptions behind these bootstrap methods. We will focus on constructing percentile intervals for the median since this parameter has not been discussed in this context as thoroughly as the mean and proportion \parencite[e.g.][]{engel2010teaching, hesterberg2015teachers, hayden2019questionable}. However, these simulations can be extended to a variety of parameters and any of the bootstrap intervals discussed earlier. To ensure accessibility, these simulations should be implemented using code that best fits the background of the students in the course. There are faster implementations that rely on \texttt{for} loops. However, asking students to apply such skills could hinder accessibility for those who are just learning to code. We provide the full \texttt{R} code needed to perform these simulations in the appendices.  

\subsection{Simulation Size and Time Constraints}

In most cases we will perform 1000 simulations with $B = 999$ because this size of simulations fits the time and computational constraints of the typical introductory statistics course. Performing $100,000$ or even 1 million simulations and using $B = 9999$ would no doubt lead to more accurate results. Additionally, if instructors and students repeat entire sets of simulations many times they can also obtain standard errors for the performance metrics that we discuss. 

However, we envision instructors performing these simulations in a timed teaching setting. Our use of a small number of simulations is meant to take time constraints into consideration. Instructors should be aware that there is a trade-off between the accuracy and ease of these simulations. It should also be noted that many methods taught in an introductory statistics course rely on some type of approximation and are not exact. This should be kept in mind when evaluating results about the accuracy of intervals.

Instructors can adapt the code provided to perform more thorough simulations outside of class while having students perform smaller scale simulations during class. After the students have obtained and discussed the results of smaller simulations, the instructor can then present the more thorough results and discuss differences and similarities. This allows students to still interact with the simulations in a timely manner without reducing the quality of the simulation results. It also allows students to see how the number of simulations performed can impact the results. 

\subsection{Simulated Data Versus Real World Data}

We use simulated data because it allows the instructor and students to evaluate each of these methods in controlled scenarios. After using simulations to discuss the assumptions of these methods, the instructor should use real data to generate plots of the bootstrap distribution and walk students through the typical application. 

Our pedagogical discussions will focus on a simulation-based explanation of these intervals which is a precursor to actual application. For example applications of the interval in single-use cases we point readers to the introductory statistics textbooks listed in Section 1. 

\subsection{Varying the Underlying Population Data}

One way to communicate the assumptions of the percentile interval to students is to simply have them construct the interval and evaluate its performance when its assumptions are or are not met. When there is severe skewness in the underlying population or the sample size is small it may be unreasonable to assume that a normalizing transformation of the sampling distribution exists. To investigate this students should first generate normal and skewed population distributions. 

The code below accomplishes this using 1 million draws from Normal($\mu = 4, \sigma^2 = 4$) and Exponential($\lambda = 1/4$) distributions. These are symmetric and right-skewed, respectively, and the skewness can be adjusted in the latter scenario through the \texttt{rate} argument. Setting the number of draws \texttt{ndraws} as a variable allows the students to propagate adjustments to this value all throughout the code in the future without changing the actual value in every place where it is used.

\begin{footnotesize}
\begin{verbatim}

# set seed for reproducibility
set.seed(9853)

## set number of draws
ndraw <- 1e+06

## draw from various populations
popdat <- data.frame(values = c(rnorm(ndraw, mean = 4, sd = 2), rexp(ndraw, rate = 0.25)),
                     pop = rep(c("Normal", "Right-Skewed"), each = ndraw))
                     
\end{verbatim}    
\end{footnotesize}

After simulating values from the populations students can then use the code below to plot these distributions side-by-side. The code below includes options for doing this with \texttt{ggplot} or \texttt{baseR}. Titles and labels can be adjusted using a \texttt{labs} layer in \texttt{ggplot} or by setting \texttt{xlab}, \texttt{ylab}, and \texttt{main} in \texttt{hist}.

\begin{footnotesize}
\begin{verbatim}

## plot with ggplot
ggplot(popdat) +
  geom_histogram(aes(values), color = "white") +
  facet_wrap(~ pop, scales = "free_x")

# OR

## plot with baseR
par(mfrow = c(1, 2))
hist(popdat$values[popdat$pop == "Normal"])
hist(popdat$values[popdat$pop == "Right-Skewed"])

\end{verbatim}    
\end{footnotesize}

\begin{figure}
    \centering
    \includegraphics[scale = 0.6]{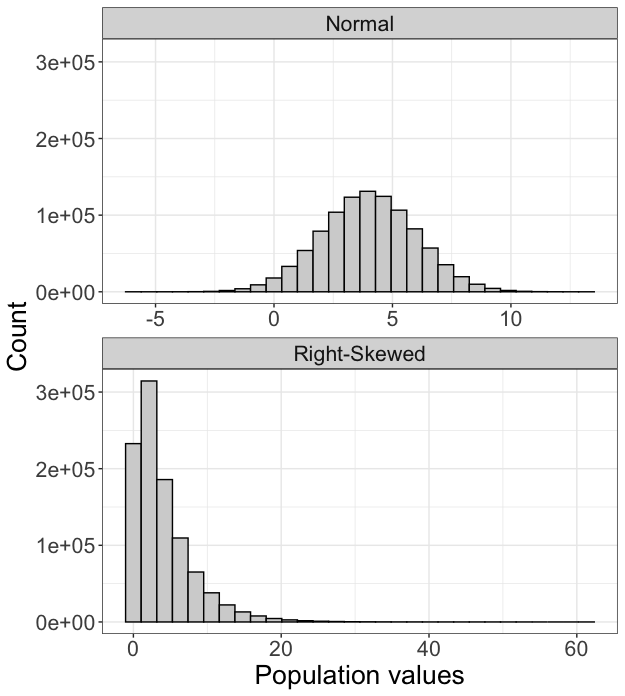}
    \caption{Normal and right-skewed population distributions that students will generate samples from. Underlying populations are Normal($\mu = 4, \sigma^2 = 4$) and Exponential($\lambda = 0.25$)}
    \label{fig:popdists}
\end{figure}

Figure \ref{fig:popdists} gives the two distributions created using \texttt{ggplot2}. Most students will clearly see the difference between the two underlying population distributions. Instructors can further prompt student understanding by asking questions such as ``What values are most likely for the Normal distribution? For the right-skewed?'' or ``What values are least likely?''

In order to make the course accessible, time should be given to students to run through the code line-by-line, to ask questions, and to make changes to the underlying population and observe the outcome. Instructors may also find it helpful to navigate students to the help pages for some of the functions used to simulate the data (e.g. \texttt{?rexp}, \texttt{?rnorm}, \texttt{?rep}). 

\subsection{Generating Sampling Distributions}

After students have explored the underlying populations, they should move forward to simulate sampling distributions from each population. Starting with a small sample size of $N = 10$ students can calculate the medians of many samples from each population. Below we first set the number of simulations \texttt{nsim} which denotes the number of samples that will be drawn from the population. Then we set the sample size \texttt{n} for each sample.

\begin{footnotesize}
\begin{verbatim}
    
# set number of simulations
nsim <- 1e+05
    
## set sample size
n <- 10

\end{verbatim}
\end{footnotesize}

The total number of values that we need to draw from each population is \texttt{n*nsim}. Therefore, students can use \texttt{rnorm} just once to generate this very large sample and then use \texttt{matrix} to organize these into columns. An example of this is given below for the normal population. Each column represents one sample from the population. Students can then use \texttt{apply} to calculate the column medians and save this as a dataset. They should repeat this for the right-skewed population.

\begin{footnotesize}
\begin{verbatim}

## generate nsim sample medians from a normal population
medsnorm <- rnorm(n*nsim, mean = 4, sd = 2) %>%
  matrix(nrow = n) %>%
  apply(2, median) %>%
  data.frame(median = ., pop = "Normal")
                    
\end{verbatim}
\end{footnotesize}

One benefit to the code above is that it does not include nested functions. Rather we elect to use the pipe (\texttt{\%>\%}) operator from \texttt{dplyr}. Though a function like \texttt{replicate} could be used with nested functions, this may be difficult for some students to read and understand at first. For example, the code below achieves the same result for both populations at once, but it is quite difficult to explain because the order of functions has to be read from right-to-left.

\begin{footnotesize}
\begin{verbatim}

samplingdists <- data.frame(medians = c(replicate(nsim, median(rnorm(n, mean = 4, sd = 2))),
                                replicate(nsim, median(rexp(n, rate = 0.25)))),
                    pop = rep(c("Normal", "Right-Skewed"), each = nsim))
                    
\end{verbatim}
\end{footnotesize}

\begin{figure}
    \centering
    \includegraphics[scale = 0.6]{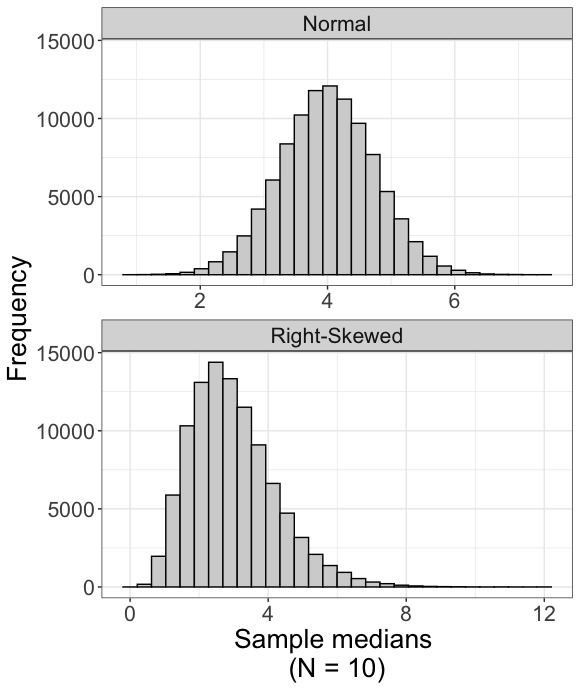}
    \caption{Distribution of sample medians where samples are size $N = 10$ and the shape of the underlying population is normal or right-skewed.}
    \label{fig:sampdistmeds10}
\end{figure}

The instructor should select the option that they deem best for their students. Once students have obtained the sampling distributions and understood the code needed to do so they should plot them. This can be achieved using code similar to that provided earlier for the population distributions. The sampling distributions when $N = 10$ are given in Figure \ref{fig:sampdistmeds10} for both population scenarios. 

Instructors can discuss the normality assumption of the percentile interval with students and walk them through how to read these images. Then they might ask questions like: ``How are the sampling distributions similar to their corresponding population distributions? How are they different?'' ``In which cases are the assumptions of the percentile interval met?'' Students can see that when the sample size is 10 the sampling distribution of the untransformed median is skewed when the population is skewed. The severity of skew in the sampling distribution depends on how skewed the population is. When the population is normal the sampling distribution is also normal.

\begin{figure}
    \centering
    \includegraphics[scale = 0.6]{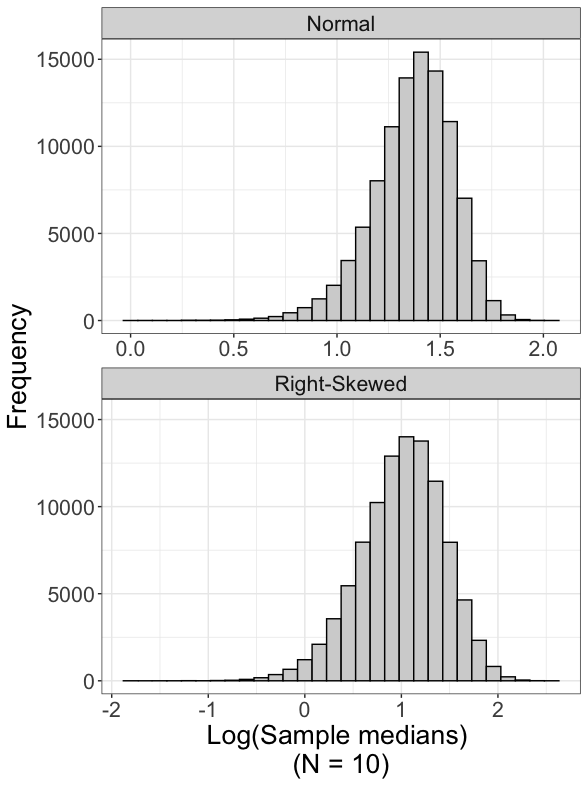}
    \caption{Distribution of log-transformed sample medians where samples are size $N = 10$ and the shape of the underlying population is normal or right-skewed.}
    \label{fig:logsampdistmeds10}
\end{figure}

However, the percentile interval just assumes that \textit{some} approximately normal transformation of the sampling distribution exists, and this does not have to be the identity transformation. To go further in learning, instructors might want to ask, ``How can we make these distributions more normal?'' Students might suggest a few transformations like the natural log. Then instructors can transform the medians for these cases and discuss the resulting sampling distributions. The log-transformed distributions in Figure \ref{fig:logsampdistmeds10} are still somewhat skewed though there are slight improvements over those in Figure \ref{fig:sampdistmeds10} for the right-skewed population.

\subsection{Varying the Sample Size}

\begin{figure}[!h]
    \centering
    \includegraphics[scale = 0.60]{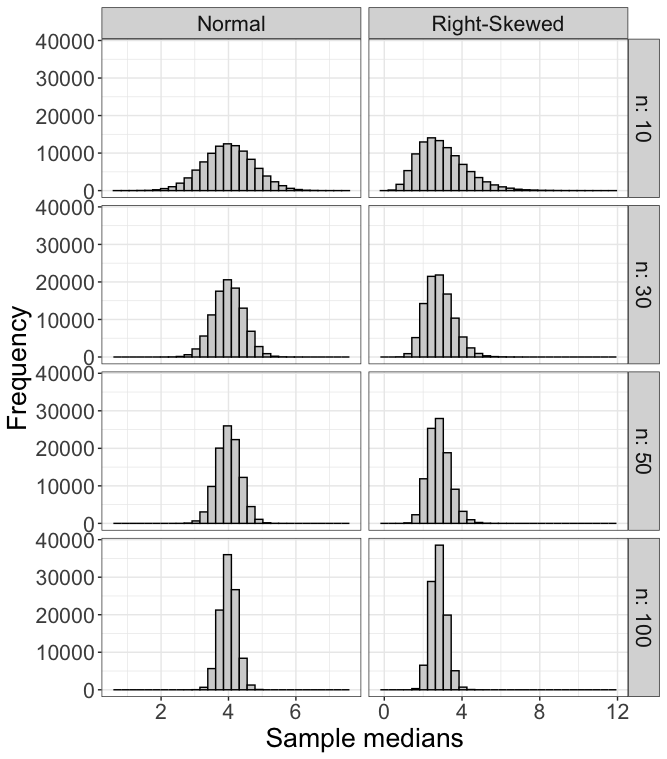}
    \caption{Sampling distributions for the median calculated using various sample sizes. Students can use plots such as these to understand whether the normality assumption of the percentile interval is met for the untransformed median. Note that the x-axes are allowed to vary to capture changes in the range of the population.}
    \label{fig:gridofmedians}
\end{figure}

Students can try other transformations but eventually the instructor should bring up the small sample size. They can ask questions like ``How might the small sample size impact these results?''. Figure \ref{fig:gridofmedians} is an updated visualization with varying sample sizes which the instructor can produce. The code is redundant and so we place it in the Appendix. Using this visualization, students can clearly see that the assumptions behind the percentile interval become more reasonable as the sample size increases.

This graphic can be used to communicate to students why we might expect the percentile interval to perform better as the sample size increases. Moreover, teachers can use this as an opportunity to explain the difference between an assumption and a condition. When the sample size is large (a \textit{condition}) it is reasonable to assume that there exists some normalizing transformation of the sampling distribution of the sample median (an \textit{assumption}).


\subsection{The Coverage Proportion of Intervals}

To help students understand what instructors mean when they say that a method ``performs better'' or is ``more accurate'' instructors can introduce them to the concept of the coverage proportion, $C$. This is the proportion of intervals that actually capture the true population parameter. For a $(1 -\alpha)^*100\%$ bootstrap interval we aim for a coverage proportion of $1 - \alpha$. Though we introduce this performance metric to students using its theoretical name, our results are simulation-based. Instructors can point out that the results students obtain are approximations and that their accuracy depends on how many simulations they perform. 

Students can use the code below to generate many 95\% percentile intervals for the population median using samples of varying sizes from various populations. The samples are organized into columns of a matrix and then \texttt{apply} is used to calculate a percentile interval from each of these. The function \texttt{percentileFUN} (see the Appendix) requires the following arguments: the vector of original sample values, the name of the statistic to calculate (e.g. mean, median, proportion), the number of bootstrap samples to obtain, the significance level, and the direction of the alternative hypothesis. The argument \texttt{statistic} will be evaluated as a function name so it should match an actual function in \texttt{R} or one that is written and saved to the global environment.

\begin{footnotesize}
\begin{verbatim}

## generate nint intervals
intervals <- rnorm(n*nint, mean = 4, sd = 2) %>%
  matrix(nrow = n) %>%
  apply(2, percentileFUN, statistic = "median", B = 999, alpha = 0.05, 
        alternative = "two.sided")
        
\end{verbatim}
\end{footnotesize}

The intervals are saved as a $2 \times 1000$ matrix where each column contains the bounds of an interval. To calculate the coverage proportion we determine whether each interval contains $\mu = 4$ as in the code below. One advantage of using this code is that student can clearly see that we are checking whether the lower bound is less than or equal to 4 and the upper bound is greater than or equal to 4. This gives students a fairly clear understanding of what ``containment of the true parameter'' actually means.

\begin{verbatim} 
apply(intervals, 2, function(x) x[1] <= 4 & x[2] >= 4) %>% mean() 
\end{verbatim} 

This querying process will return \texttt{TRUE} or \texttt{FALSE} for each interval. We pipe this result into \texttt{mean} which will return the proportion of \texttt{TRUE} results. Alternatively students can save the results of \texttt{apply} to an object, say \texttt{containmentvec}, and use one of the two options below. \begin{verbatim} sum(containmentvec == TRUE)/length(containmentvec)\end{verbatim} \begin{verbatim} mean(containmentvec)
\end{verbatim} The first option may be more beneficial to some students since they can clearly see how the proportion is calculated from a set of logical values. 


After understanding the code, students can perform additional simulations for each of the sample sizes and populations in Figure \ref{fig:gridofmedians}. If performing all eight simulations is infeasible then instructors can generate the results outside of class time. However, instructors should make sure that students grasp \textit{why} they are changing the population and sample size. The purpose is to understand how the performance of the percentile interval changes as the underlying population and sample size change. Table \ref{tab:covpropsupsmall} contains the coverage proportions for all eight scenarios.

\begin{table}[!h]
    \centering
    \begin{tabular}{c|c|c}
    \hline
     & Normal & Right-skewed \\
     \hline
    N = 10 & 0.936 & 0.934 \\
    N = 30 & 0.939 & 0.975 \\
    N = 50 & 0.948 & 0.934 \\
    N = 100 & 0.94 & 0.944 \\
    \hline
    \end{tabular}
    \caption{Coverage proportions of 1000 two-sided 95\% percentile intervals for the median. Samples of size N were drawn from the underlying populations. The number of bootstrap samples was 999. Students can perform simulations such as these to learn how to assess the performance of the percentile interval in different scenarios.}
    \label{tab:covpropsupsmall}
\end{table}

The coverage proportions fall below the nominal 0.95 level in most cases, especially when $N = 10$. The instructor can remind students that when $N = 10$ the identity transformation also least satisfied the assumptions of the percentile interval. Generating these results can help students to connect the conditions of the interval to its performance, however, instructors should mention that the low number of simulations and lack of replication impacts the accuracy of students' results. 

\subsection{Obtaining Standard Errors}

\begin{figure}
    \centering
    \includegraphics[scale = 0.5]{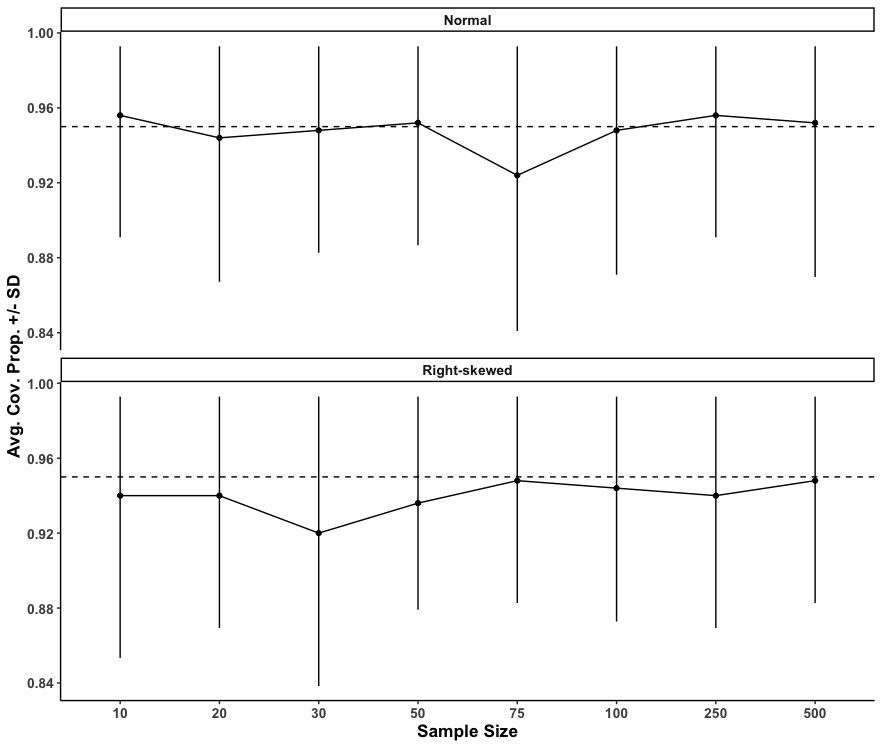}
    \caption{Average and standard deviation of 25 coverage proportions each coming from 10 95\% two-sided percentile intervals for the median using $B = 999$ bootstrap samples.}
    \label{fig:avgcovprobsd}
\end{figure}

Instructors can present the results of a larger simulation that uses more intervals, bootstrap samples, and replications of the entire process. Figure \ref{fig:avgcovprobsd} provides the average coverage proportions and their standard deviations. This was calculated out of 25 iterations of the process outlined in the previous subsection for various sample sizes using 10 bootstrap intervals. The code for this larger simulation (see the Appendix) will likely be too complex for introductory statistics students to fully understand unless if they have prior experience with computer programming. Instructors can perform these simulations ahead of time and discuss the results after students have performed the smaller simulations. 

In Figure \ref{fig:avgcovprobsd} we see that the average of the coverage proportions reaches the nominal 0.95 level denoted by the dashed line more often and more closely for the normal population. For smaller sample sizes the intervals constructed on right-skewed data consistently under-perform. However, there is still some departures from what we would expect to see in theory. For example, the average coverage proportion when $N = 75$ is lower than when $N = 10$ in the normal case. This may be due to the low number of repetitions. The instructor should set the simulation size to fit their computational resources. We again note that these simulation results do not rule out the existence of a normalizing transformation when the sample size is small but they \textit{can} help students to understand which factors impact the performance of these method and how.

\subsection{Significance Level and Power}

Instructors can also use the values plotted in Figure \ref{fig:avgcovprobsd} to discuss the Type I Error rate of this method in a variety of scenarios. Subtracting the coverage proportion from 1 will also produce an estimated significance level for the bootstrap hypothesis tests discussed in Section 3 since these are based on inverting the intervals used to calculate the coverage proportions. In this case, the nominal level would be $\alpha = 0.05$ since we constructed 95\% confidence intervals.

Additionally, calculating the proportion of intervals that \textit{do not} contain a value that is different from the true population parameter provides an estimate of the power since this is the rate at which we would \textit{reject} this false value. Students may benefit from viewing the power as a function of the hypothesized value with a place marker for the true parameter. These results could help students to identify when a bootstrap hypothesis test is conservative - that is, when its rejection rate remains low unless if the hypothesized value is quite far from the true parameter. 

Meanwhile, if the power of the test does not converge to one as the sample size increases, or if it does so more slowly in a given scenario, then the test is said to have low power. Therefore, viewing power results as a function of sample size may further help student understanding. The code for generating results for the power is more complicated than we should expect the typical introductory statistics student to understand. Instructors can perform these simulations outside of class and explain the results to students. This code is also included in the Appendix, but we omit a discussion of its results for brevity.

\subsection{Further Extensions}

Instructors can also extend these simulations to two-sample problems. Constructing and assessing percentile intervals for two-sample statistics follows a very similar process as that already outlined. Instructors may want to introduce the use of color-blind friendly palettes for visualizing population distributions by group to keep this lesson accessible to all students.

The $t$-interval with bootstrap standard error uses quantiles from a $t$-distribution and the bootstrap standard error. When the sampling distribution of the statistic is not actually $t$-distributed the studentized interval is a better alternative. The simulations that we have performed can be extended in order to communicate the assumptions behind this interval to students as well. We include code for generating studentized bootstrap intervals in the Appendix. 

By performing simulations like those that we have outlined, students can better understand the assumptions behind these bootstrap methods. As with many statistical methods, poor performance can be expected when the sample size is small and the underlying population is skewed. Checking the bootstrap distribution for approximate normality is a reliable way to determine if the percentile interval is appropriate. Instructors should also remind students that any bootstrap method can fail if the empirical CDF ($\hat{F}$) is a poor estimate of the population CDF ($F$). When the sample size is small students should be cautious of using the bootstrap since we risk trying to estimate a population with a small amount of information. 

\section{Conclusions}

In this article, the theoretical details of the $t$-interval with bootstrap standard error, percentile, and studentized intervals and their corresponding hypothesis tests were discussed. We showed that these methods have important underlying assumptions and discussed simulations that can be used to present these details to students with active and visual learning. We also encouraged instructors to take advantage of the simulation setting to discuss performance metrics such as the coverage proportion, significance level, and power. 

The use of simulations and other statistical computing methods in the classroom equips students with a variety of tools to use in many situations. It also increases students' retention of concepts and aids the teacher in explaining complex topics. With this article, we aim to benefit both teacher and student by presenting novel ideas for communication the assumptions behind simple bootstrap methods to students in an accessible manner. It is important that each student understand the usefulness and the correct scope of these methods before leaving the classroom so that they are equally equipped to handle a variety of situations. 

\section*{Disclosure Statement}

The authors report there are no competing interests to declare.

\appendix

\footnotesize

\section{Appendix}

\subsection{Obtain population distributions}

\begin{verbatim}

# load needed packages
library(tidyverse)

#################################
#################################

# set seed for reproducibility
set.seed(9853)

# generate approximate population distributions

## set number of draws
ndraw <- 1e+06

## draw from various populations
popdat <- data.frame(values = c(rnorm(ndraw, mean = 4, sd = 2), rexp(ndraw, rate = 0.25)),
                     pop = rep(c("Normal", "Right-Skewed"), each = ndraw))

## plot distributions
ggplot(popdat) +
  geom_histogram(aes(values)) +
  facet_wrap(~ pop, scales = "free_x")

## plot with baseR
par(mfrow = c(1, 2))
hist(popdat$values[popdat$pop == "Normal"])
hist(popdat$values[popdat$pop == "Right-Skewed"])

\end{verbatim}

\subsection{Obtain sampling distributions}

\begin{verbatim}

# set number of simulations
nsim <- 1e+05

# obtain sample medians from different populations

## set sample size
n <- 10

## NESTING: generate sampling distribution for right-skewed data
set.seed(456)
samplingdists <- data.frame(median = c(replicate(nsim, median(rnorm(n, mean = 4, sd = 2))),
                                replicate(nsim, median(rexp(n, rate = 0.25)))),
                    pop = rep(c("Normal", "Right-Skewed"), each = nsim))

## PIPING: generate sampling distribution for right-skewed data
set.seed(456)
medsnorm <- rnorm(n*nsim, mean = 4, sd = 2) %>%
  matrix(nrow = n) %>%
  apply(2, median) %>%
  data.frame(median = ., pop = "Normal")

## PIPING: generate sampling distribution for right-skewed data
set.seed(456)
medsexp <- rexp(n*nsim, rate = 0.25) %>%
  matrix(nrow = n) %>%
  apply(2, median) %>%
  data.frame(median = ., pop = "Right-Skewed") 

## rbind all data together
samplingdists <- rbind(medsnorm, medsexp)

## plot distributions
ggplot(samplingdists) +
  geom_histogram(aes(median)) +
  facet_wrap(~ pop, scales = "free_x") +
  labs(x = "Sample medians \n(N = 10)")

## LOG TRANSFORM plot distributions w/ more niceties for article
ggplot(samplingdists) +
  geom_histogram(aes(log(median)), color = "black", fill = "lightgray") +
  facet_wrap(~ pop, scales = "free_x", nrow = 2) +
  labs(x = "Log(Sample medians) \n(N = 10)", y = "Frequency") +
  theme_bw() +
  theme(text = element_text(size = 20))

\end{verbatim}

\subsection{Varying the sample sizes}

\begin{verbatim}

# generate sampling distributions for varying n
## set seed for reproducibility
set.seed(456)
##
n <- 30
samplingdists30 <- data.frame(median = c(replicate(nsim, median(rnorm(n, mean = 4, sd = 2))),
                                          replicate(nsim, median(rexp(n, rate = 0.25)))),
                              pop = rep(c("Normal", "Right-Skewed"), each = nsim),
                              n = n)
##
n <- 50
samplingdists50 <- data.frame(median = c(replicate(nsim, median(rnorm(n, mean = 4, sd = 2))),
                                          replicate(nsim, median(rexp(n, rate = 0.25)))),
                              pop = rep(c("Normal", "Right-Skewed"), each = nsim),
                              n = n)
##
n <- 100
samplingdists100 <- data.frame(median = c(replicate(nsim, median(rnorm(n, mean = 4, sd = 2))),
                                          replicate(nsim, median(rexp(n, rate = 0.25)))),
                              pop = rep(c("Normal", "Right-Skewed"), each = nsim),
                              n = n)

# combine data
samplingdistsALL <- rbind(samplingdists,
                          samplingdists30, 
                          samplingdists50,
                          samplingdists100)

# plot sampling distributions for varying sample sizes
ggplot(samplingdistsALL) +
  geom_histogram(aes(median), color = "black", fill = "lightgray") +
  facet_grid(n ~ pop, scales = "free_x",
             labeller = labeller(.rows = label_both)) +
  labs(x = "Sample medians", y = "Frequency") +
  theme_bw() +
  theme(text = element_text(size = 20))

\end{verbatim}

\subsection{Function for the percentile interval}

\begin{verbatim}

# function for calculating percentile interval
percentileFUN <- function(origsample, statistic, B, alpha, alternative){
  ## create matrix where each column is a bootstrap sample
  b <- matrix(sample(origsample, replace = TRUE, size = length(origsample)*B), ncol = B)
  ## calculate bootstrap sample statistics
  stats <- apply(b, 2, function(x) match.fun(statistic)(x))
  ## return appropriate interval
  if(alternative == "two.sided"){return(c(sort(stats)[(B+1)*(alpha/2)], 
                                          sort(stats)[(B +1 )*(1-(alpha/2))]))}
  if(alternative == "less"){return(c(-Inf, sort(stats)[(B+1)*(1-alpha)]))}
  if(alternative == "greater"){return(c(sort(stats)[(B+1)*(alpha)], Inf))}
}

\end{verbatim}

\subsection{Obtain coverage proportions}

\begin{verbatim}

# calculate coverage proportions for percentile interval for various sample sizes and populations

## set seed for reproducibility
set.seed(456)

## set number of intervals
nint <- 1000

## set sample size - CHANGE THIS
n <- 10

## NORMAL POPULATION

## generate nint intervals
intervals <- rnorm(n*nint, mean = 4, sd = 2) %>%
  matrix(nrow = n) %>%
  apply(2, percentileFUN, statistic = "median", B = 999, alpha = 0.05, 
        alternative = "two.sided")

## calculate coverage proportions (each column gives interval bounds)
apply(intervals, 2, function(x) x[1] <= 4 & x[2] >= 4) %>% mean()

## EXPONENTIAL POPULATION

## generate nint intervals
intervals <- rexp(n*nint, rate = 0.25) %>%
  matrix(nrow = n) %>%
  apply(2, percentileFUN, statistic = "median", B = 999, alpha = 0.05, 
        alternative = "two.sided")

## calculate coverage proportions (each column gives interval bounds)
truemedian <- log(2)/0.25
apply(intervals, 2, function(x) x[1] <= truemedian & x[2] >= truemedian) %>% mean()

## other way to calculate coverage proportion to see the math more clearly
containmentvec <- apply(intervals, 2, function(x) x[1] <= 4 & x[2] >= 4)
sum(containmentvec == TRUE)/length(containmentvec)
mean(containmentvec)

\end{verbatim}

\subsection{Obtain standard errors for coverage proportions}

\begin{verbatim}

# sample sizes
sampsizes <- c(10, 20, 30, 50, 75, 100, 250, 500)

# code to evaluate samples of each size from each population
coderun <- c(paste("rnorm(", sampsizes, ", mean = 4, sd = 2)", sep = ""),
             paste("rexp(", sampsizes, ", rate = 0.25)", sep = ""))

# save each population
pops <- rep(c("Normal", "Right-skewed"), each = length(sampsizes))

# parameter to check for containment of
parcheck <- rep(c(4, log(2)/0.25), each = length(sampsizes))


# list to save intervals to
interval_list <- list()

for(i in 1:25){ # number of coverage proportions to generate
  
  interval_list[[i]] <- numeric(length(coderun))
  
  for(j in 1:length(coderun)){ # for each sample size and population combination
    
    # set seed to use same sample and bootstrap samples for each iteration over (i,j)
    set.seed(j+((i-1)*32))
    # generate many intervals
    interval_list[[i]][[j]] <- replicate(10, 
                    percentileFUN(origsample = eval(parse(text = coderun[j])),
                                                           statistic = "median", 
                                                           B = 999, 
                                                           alpha = 0.05, 
                                                           alternative = "two.sided")) %>%
      apply(2, function(x) x[1] <= parcheck[j] & x[2] >= parcheck[j]) %>% 
      mean()
    
  }
  

}

# read in results from simulations
interval_dat <- do.call("cbind", interval_list) %>%
  apply(1, function(x) c(mean = mean(x), sd = sd(x))) %>%
  data.frame() %>%
  t() %>%
  cbind(sampsizes) %>%
  data.frame() %>%
  cbind(pops = pops)

# plot it
interval_dat %>%
  ggplot() +
  geom_line(aes(as.factor(sampsizes), mean, group = as.factor(pops))) +
  facet_wrap(~ pops) +
  geom_point(aes(as.factor(sampsizes), mean)) +
  geom_hline(aes(yintercept = 0.95), linetype = "dashed") +
  geom_linerange(aes(as.factor(sampsizes), ymin = mean - sd, ymax = min(mean + sd, 1))) +
  labs(y = "Avg. Cov. Prop. +/- SD", x = "Sample Size") +
  theme_classic() +
  theme(legend.position = "none", text = element_text(face = "bold", size = 13))

\end{verbatim}

\subsection{Hypothesis testing and power extensions}

\begin{verbatim}

## set seed for reproducibility
set.seed(234509)

## generate many intervals
n <- 10
intervals_10 <- replicate(1000, percentileFUN(origsample = rnorm(n, mean = 4, sd = 2),
                                              statistic = "median", 
                                              B = 999, 
                                              alpha = 0.05, 
                                              alternative = "two.sided"))
## generate many intervals
n <- 30
intervals_30 <- replicate(1000, percentileFUN(origsample = rnorm(n, mean = 4, sd = 2),
                                              statistic = "median", 
                                              B = 999, 
                                              alpha = 0.05, 
                                              alternative = "two.sided"))
## generate many intervals
n <- 100
intervals_100 <- replicate(1000, percentileFUN(origsample = rnorm(n, mean = 4, sd = 2),
                                               statistic = "median", 
                                               B = 999, 
                                               alpha = 0.05, 
                                               alternative = "two.sided"))

## generate many intervals
n <- 10
intervals_10_exp <- replicate(1000, percentileFUN(origsample = rexp(n, rate = 0.25),
                                                  statistic = "median", 
                                                  B = 999, 
                                                  alpha = 0.05, 
                                                  alternative = "two.sided"))
## generate many intervals
n <- 30
intervals_30_exp <- replicate(1000, percentileFUN(origsample = rexp(n, rate = 0.25),
                                                  statistic = "median", 
                                                  B = 999, 
                                                  alpha = 0.05, 
                                                  alternative = "two.sided"))
## generate many intervals
n <- 100
intervals_100_exp <- replicate(1000, percentileFUN(origsample = rexp(n, rate = 0.25),
                                                   statistic = "median", 
                                                   B = 999, 
                                                   alpha = 0.05, 
                                                   alternative = "two.sided"))

# plot to see power of hypothesis test for each population sample size combination
# first organize the data
data.frame(
  
  containprop = c(
    
    ## what proportion of intervals contain 0? 1? 2? ... 9? 
    lapply(list(0, 1, 2, 3, 4, 5, 6, 7, 8, 9), 
           function(y) mean(apply(intervals_10, 2, function(x) x[1] <= y & x[2] >= y))) %>%
      unlist(),
    
    lapply(list(0, 1, 2, 3, 4, 5, 6, 7, 8, 9), 
           function(y) mean(apply(intervals_30, 2, function(x) x[1] <= y & x[2] >= y))) %>%
      unlist(),
    
    lapply(list(0, 1, 2, 3, 4, 5, 6, 7, 8, 9), 
           function(y) mean(apply(intervals_100, 2, function(x) x[1] <= y & x[2] >= y))) %>%
      unlist(),
    
    lapply(list(0, 1, 2, log(2)/0.25, 4, 5, 6, 7, 8, 9), 
           function(y) mean(apply(intervals_10_exp, 2, function(x) x[1] <= y & x[2] >= y))) %>%
      unlist(),
    
    lapply(list(0, 1, 2, log(2)/0.25, 4, 5, 6, 7, 8, 9), 
           function(y) mean(apply(intervals_30_exp, 2, function(x) x[1] <= y & x[2] >= y))) %>%
      unlist(),
    
    lapply(list(0, 1, 2, log(2)/0.25, 4, 5, 6, 7, 8, 9), 
           function(y) mean(apply(intervals_100_exp, 2, function(x) x[1] <= y & x[2] >= y))) %>%
      unlist()
    
    
  ),
  
  ## values that each coverage proportion was calculated for
  theta = c(0:9, 0:9, 0:9, 0:2, log(2)/0.25, 4:9, 0:2, log(2)/0.25, 4:9, 0:2, log(2)/0.25, 4:9),
  
  ## sample size corresponding to coverage proportion calculated
  sampsize = rep(c(10, 30, 100, 10, 30, 100), each = 10),
  
  ## population that underlying samples of intervals was from
  pop = c(rep("Normal \npopulation", 30), rep("Right-Skewed \npopulation", 30))
  
  
) %>%
  ggplot() +
  geom_line(aes(theta, 1 - containprop, linetype = pop)) +
  geom_vline(data = data.frame(xintercept = c(4, log(2)/0.25), 
                               pop = c("Normal \npopulation", "Right-Skewed \npopulation")),
             mapping = aes(xintercept = xintercept, linetype = pop)) +
  geom_text(data = data.frame(xintercept = c(4, log(2)/0.25), 
                              pop = c("Normal \npopulation", "Right-Skewed \npopulation")),
            mapping = aes(x = xintercept - 0.3, y = 0.60, label = "True \nvalue")) +
  facet_wrap(~ sampsize, nrow = 3, 
             labeller = as_labeller(c("10" = "N = 10", "30" = "N = 30", "100" = "N = 100"))) +
  scale_linetype_discrete("") +
  labs(x = "Hypothesized value", y = "Rejection rate") +
  theme_bw() +
  theme(legend.position = "bottom",
        text = element_text(size = 15, face = "bold"))

\end{verbatim}

\subsection{Function for the studentized interval}

\begin{verbatim}


studentized <- function(sample, parameter, B = 999, siglevel = 0.05, onlyint = FALSE, M = 25){
  
  # CREATE INITIAL BOOTSTRAP OBJECT
  
  # stopping clause to make sure the arguments are of the correct structure
  # stopping clause to make sure the arguments are of the correct structure
  stopifnot(is.vector(sample))
  stopifnot(!is.list(sample))
  stopifnot(!is.expression(sample))
  stopifnot(is.character(parameter))
  stopifnot(exists(parameter))
  stopifnot(length(match.fun(parameter)(sample)) == 1)
  stopifnot(as.integer(B) == B)
  stopifnot(B <= 5000)
  stopifnot(siglevel > 0 & siglevel < 1)
  stopifnot(onlyint == FALSE | onlyint == TRUE)
  stopifnot(M > 15 & M < 50)
  
  # drop NA's from original sample before bootstrapping
  sampleuse <- stats::na.omit(sample)
  
  # resample from the original sample (w/o NA's) 'B*length(sample)' times 
  # with replacement and then organize into a matrix with B columns
  # so each column is a bootstrap sample
  b <- matrix(sample(sampleuse, replace = TRUE, size = length(sampleuse)*B), ncol = B)
  
  # calculate the bootstrap sample statistics
  stats <- numeric(B)
  
  for(i in 1:B){
    stats[i] <- match.fun(parameter)(b[,i])
  }
  
  # calculate original sample statistic
  ogstat <- match.fun(parameter)(sampleuse)
  
  # IMPLEMENT STUDENTIZED METHOD
  
  # estimate for variance of sample statistic
  uppervar <- (1/(B - 1))*sum((stats - mean(stats))^2)
  
  # estimate for variance of each bootstrap sample statistic using second level bootstrap
  lowervar <- numeric(B)
  for(i in 1:B){
    # sample from each bootstrap sample M times and obtain M second level bootstrap sample statistics
    slb <- matrix(sample(b[,i], replace = TRUE, size = length(b[,i])*M), ncol = M)
    slstats <- numeric(M)
    for(j in 1:M){
      slstats[j] <- match.fun(parameter)(slb[,j])
    }
    slogstat <- stats[i]
    # calculate variance of second level bootstrap sample statistics
    lowervar[i] <- (1/(M-1))*sum((slstats - mean(slstats))^2)
  }
  
  # distribution of mock z-statistics
  z <- (stats - ogstat)/sqrt(lowervar)
  
  # the interval
  interval <- c(ogstat - (sqrt(uppervar)*sort(z)[round((B + 1)*(1 - siglevel/2))]),
                ogstat - (sqrt(uppervar)*sort(z)[round((B + 1)*(siglevel/2))]))
  
  
  # OUTPUT
  
  if(onlyint == TRUE){
    return(interval)
  }else{
    
    # return a histogram of bootstrap sample statistics
    hist(stats, main = paste("Distribution of bootstrap \nsample statistics:", parameter),
         xlab = paste("bootstrap", parameter))
    abline(v = ogstat, col = "red", lwd = 3)
    
    # plots to check conditions
    hist(z, main = "Studentized bootstrap distribution",
         xlab = paste("studentized", parameter), pch = 19)
    
    # plot to check variance stabilized condition
    plot(stats, sqrt(lowervar), main = "Bootstrap statistic vs estimated standard error 
    \n(based on second-level bootstrap)", xlab = paste("bootstrap sample", parameter, sep = " "),
         ylab = paste("estimated SE of", parameter))
    
    # print results
    cat("The studentized bootstrap interval for the", parameter, "is: ")
    cat("(", interval[1], ", ", interval[2], ").\n", sep = "")
    cat("\nIf it is reasonable to assume that the studentized sampling distribution 
    of the \nstatistic of interest does not depend on any unknown parameters, 
    \nthen this method can be used.")
    
  }
  
}

\end{verbatim}

\newpage

\printbibliography

\end{document}